\documentclass{INTERSPEECH2023}

\interspeechcameraready

\usepackage{color, colortbl}
\usepackage{multirow}
\definecolor{Gray}{gray}{0.9}

\title{Personalized Adaptation with Pre-trained Speech Encoders 
 \\ for Continuous Emotion Recognition}
\name{Minh Tran$^*$, Yufeng Yin$^*$, Mohammad Soleymani}
\address{Institute for Creative Technologies, University of Southern California, USA}
\email{\{mtran, yin, soleymani\}@ict.usc.edu}
\begin{document}

\maketitle
\def\thefootnote{*}\footnotetext{Equal contribution}\def\thefootnote{\arabic{footnote}}

\begin{abstract} 
There are individual differences in expressive behaviors driven by cultural norms and personality. This between-person variation can result in reduced emotion recognition performance. Therefore, personalization is an important step in improving the generalization and robustness of speech emotion recognition. In this paper, to achieve unsupervised personalized emotion recognition, we first pre-train an encoder with learnable speaker embeddings in a self-supervised manner to learn robust speech representations conditioned on speakers. Second, we propose an unsupervised method to compensate for the label distribution shifts by finding similar speakers and leveraging their label distributions from the training set. Extensive experimental results on the MSP-Podcast corpus indicate that our method consistently outperforms strong personalization baselines and achieves state-of-the-art performance for valence estimation.
\end{abstract}

\section{Introduction}
With the ubiquity of voice assistive technologies, speech emotion recognition (SER) is becoming increasingly important as it allows for a more natural and intuitive interaction between humans and machines. Although SER technology has made significant progress in recent years, accurately detecting emotions from speech remains a challenging task. This is partly due to the vast variability in how people express their feelings through speech, which can depend on culture \cite{von2021cultural}, gender \cite{kring1998sex}, or age \cite{montepare2014younger}, among others. Personalization is a promising solution to address the variability of emotional expression in speech. By tailoring emotion recognition systems to match individuals' unique expressive behaviors, the approach can lead to a more robust and inclusive model that is better equipped to accurately detect emotions for a wide range of users. 

Existing studies on personalized emotion recognition generally use hand-crafted features of speech on datasets with a small number of speakers (ten or fewer speakers) \cite{chen2020two, kim2016multistage, jia2021two, vryzas2018speech, bang2018adaptive}.  Recently, SER systems achieve state-of-the-art results \cite{srinivasan2022representation, chen2021exploring} via fine-tuning large pre-trained speech encoders such as HuBERT \cite{hsu2021hubert} or wav2vec2.0 \cite{baevski2020wav2vec}. This raises three important questions: (1) What happens to the personalization gap as the number of speakers increases for fine-tuned encoders? (2) How do existing personalization methods behave when the input speech features are not fixed? (3) How can we incorporate personalization with pre-trained encoders to boost performance?

\begin{figure*}[t]
\centering
\includegraphics[width=0.86\linewidth,trim={0pt 275pt 180pt 0pt},clip]{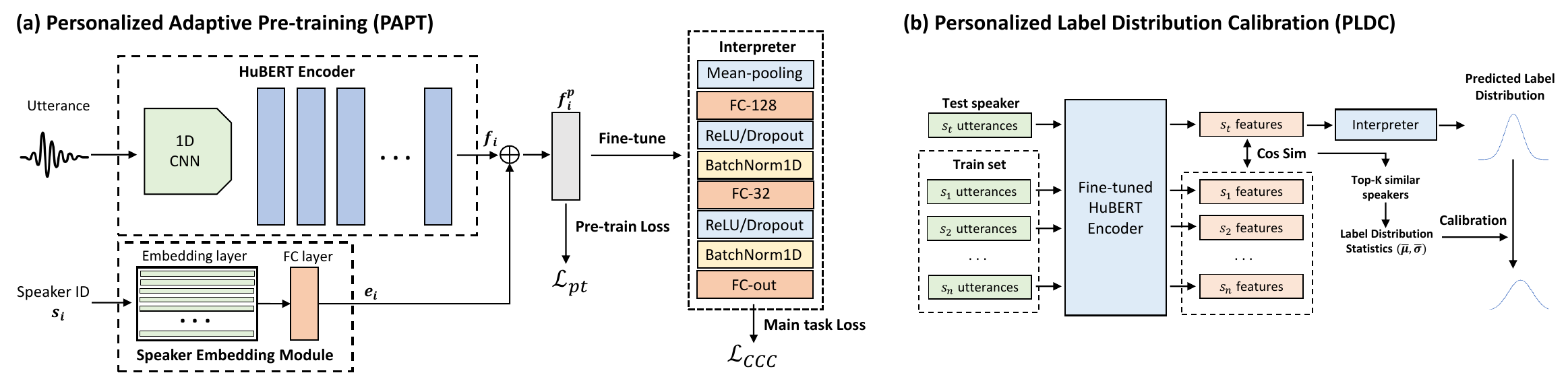}
\caption{Overview of our proposed method. (a) Personalized Adaptive Pre-Training (PAPT) pre-trains the HuBERT encoder with learnable speaker embeddings in a self-supervised manner. (b) Personalized Label Distribution Calibration (PLDC) finds similar training speakers and calibrates the predicted label distribution with the training label statistics.}
\label{fig:overview}
\end{figure*}

In this paper, we perform extensive experiments on the MSP-Podcast corpus \cite{lotfian2017building} with more than 1,000 speakers to answer these questions. We first show that as the number of speakers increases, the personalization gap (the performance difference between speaker-dependent and speaker-independent) of fine-tuned models decreases, which motivates the need for methods that adapts the pre-trained weights for personalization prior to fine-tuning. Hence, we propose to continue the pre-training process of the speech encoder jointly with speaker embeddings (see Figure \ref{fig:overview} (a)). We also introduce a simple yet effective unsupervised personalized calibration step to adjust label distribution per speaker for better accuracy (see Figure \ref{fig:overview} (b)). The proposed methods are unsupervised, requiring no prior knowledge of the test labels.
 Experimental results on arousal and valence estimation show that the proposed methods achieve state-of-the-art results for valence estimation while consistently outperforming the encoder fine-tuning baseline and a recent personalization method evaluated on the same dataset \cite{sridhar2022unsupervised}. The major contributions of this work are as follows.
(1) We propose a method for personalized adaptive pre-training to adjust the existing speech encoders for a fixed set of speakers. (2) We propose an unsupervised personalized post-inference technique to adjust the label distributions. 
(3) We provide extensive experimental results along with an ablation study to demonstrate the effectiveness of the methods.
(4) We further show that our methods can be extended to unseen speakers without the need to re-train any component, achieving superior performance compared to the baselines. 
\section{Related work}
\noindent \textbf{Adaptive Pre-training.} Adaptive pre-training resumes the pre-training process for the pre-trained encoders on either domain data (domain adaptive pre-training) or task data (task adaptive pre-training) to improve the downstream task performance of a specific domain or dataset \cite{gururangan2020don}. The method has been shown to be highly effective for a wide range of natural language processing and computer vision applications such as machine translation \cite{rubino2020intermediate}, sentiment analysis \cite{xu2019bert}, and image classification \cite{kim2022broad}. In the field of speech emotion recognition, Chen \textit{et al.} \cite{chen2021exploring} propose a novel pseudo-label generation method in combination with task-adaptive pre-training for wav2vec2.0 \cite{baevski2020wav2vec} to boost emotion recognition accuracy. However, there is no prior work exploring personalized adaptive pre-training.

\noindent \textbf{Personalized Speech Emotion Recognition.} There have been a few methods \cite{kim2016multistage, bang2018adaptive, chen2020two, jia2021two, sridhar2022unsupervised} proposed for personalized SER. However, most of the existing work are validated on datasets with limited speakers.
Most relevant to our work is the unsupervised personalized method proposed by Sridhar \textit{et al.} \cite{sridhar2022unsupervised}, which is validated on the same dataset (MSP-Podcast) as in this paper. They propose to find speakers in the train set to form the adaptation set whose acoustic patterns closely resemble those of the speakers in the test set. Specifically, they apply Principal Component Analysis (PCA) on the feature set proposed for the computational paralinguistics challenge (ComParE) \cite{schuller2013interspeech} and fit Gaussian Mixture Models to measure the speaker similarity based on the KL divergence metric. Samples of the selected speakers are given more weight during the training process. With a light architecture, \textit{i.e.}, the multi-layer perceptron, the method is shown to be highly effective for valence personalization. However, it requires extra training (model adaptation) during inference and thus can not be extended to new speakers.

In contrast to prior work, we explore personalization with fine-tuned encoders instead of pre-extracted features, which achieves superior performance compared to the best-performing models. For example, our weakest baseline (HuBERT-large fine-tuning) achieves a two times higher Concordance Correlation Coefficient (CCC) compared to the reported results from Sridhar \textit{et al.} \cite{sridhar2022unsupervised} for valence estimation. More importantly, our method is extensible and remains effective for unseen speakers without the need to re-train any components.

\section{Preliminary information}
\noindent \textbf{Problem Formulation.} Unsupervised personalized speech emotion recognition: Given a speech dataset containing $N$ utterances with emotion labels (arousal or valence) and speaker IDs $\mathcal{D} = \{(u_i, y_i, s_i)\}^N_{i=1}$. 
We assume access to all information except for the emotion labels of the test set during the training phase. Our goal is to produce a robust emotion recognition model that performs better than a model exposed to the same amount of data excluding speaker ID information. We further want our method to be extensible to new speakers outside of $\mathcal{D}$.


\noindent \textbf{Dataset.} We use the MSP-Podcast corpus \cite{lotfian2017building} as our dataset $\mathcal{D}$. MSP-Podcast is the largest corpus for speech emotion recognition in English, containing emotionally rich podcast segments retrieved from audio-sharing websites. Each utterance in the dataset is annotated using crowd-sourcing with continuous labels of arousal, valence, and dominance along with categorical emotions. In this paper, we focus on arousal and valence estimation. The labels range from 1 to 7. The dataset contains pre-defined train, validation, and test sets, namely $\mathcal{D}_{tr}$, $\mathcal{D}_{val}$, $\mathcal{D}_{te}$, which are subject independent. We use two versions of the dataset, namely v1.6 and v1.10, for the experiments. To be consistent with prior studies \cite{lin2021chunk, sridhar2022unsupervised, srinivasan2022representation}, most of our experiments are based on MSP-Podcast v1.6. We remove all the utterances marked with ``Unknown" speakers in accordance with our problem formulation. Following Sridhar \textit{et al.} \cite{sridhar2022unsupervised}, we split the test set into two subsets \textit{test-a} and \textit{test-b} that share the same set of speakers. Each speaker in \textit{test-a} contains 200s of speech in total while \textit{test-b} contains the rest of the recordings. \textit{test-a} is used to train speaker-dependent models along with the train set. For experiments on the unseen speakers, we evaluate the models on the speakers who are in the v1.10 test set but not in the v1.6 test set, namely \textit{test-c}. Table \ref{tab:statistics} provides the details and statistics of our splits for the MSP-Podcast dataset.

\begin{table}[t]
\footnotesize
\caption{Details and statistics of our splits for MSP-Podcast.}
\centering
\scalebox{0.9}{\begin{tabular}{l|ccccc}
\toprule
\rowcolor{Gray}
split & train & validation & test-a & test-b & test-c \\
\midrule
\# utterances & 26470 & 5933 & 1684 & 8434 & 7304 \\
\# speakers & 987 & 41 & 50 & 50 & 62 \\
total duration & 44.2h & 10h & 2.9h & 13.8h & 9.5h \\
corpus version & v1.6 & v1.6 & v1.6 & v1.6 & v1.10 \\
\bottomrule
\end{tabular}}
\label{tab:statistics}
\end{table}

\noindent \textbf{Pre-trained Speech Encoder.} In this work, we use HuBERT \cite{hsu2021hubert} as our pre-trained encoder $\text{E}$ due to its superior performance \cite{srinivasan2022representation, wagner2022dawn}. HuBERT consists of two main components, namely a 1D CNN and a Transformer encoder \cite{vaswani2017attention}. The 1D CNN takes raw waveforms as inputs and returns low-level feature representations of speech. Then, the features are passed into the Transformer encoder to generate high-level feature representations via the self-attention mechanism. During the pre-training process, HuBERT first generates pseudo-labels by performing K-means clustering on the pre-extracted features, \textit{e.g.}, MFCCs. Then, the model learns in a self-supervised manner through the task of predicting pseudo-labels for randomly masked frames. Therefore, the pre-training loss $\mathcal{L}_{pt}$ for HuBERT can be defined as the sum of the cross-entropy loss computed over the masked frames.

\begin{figure}[t]
\centering
\includegraphics[width=0.75\linewidth]{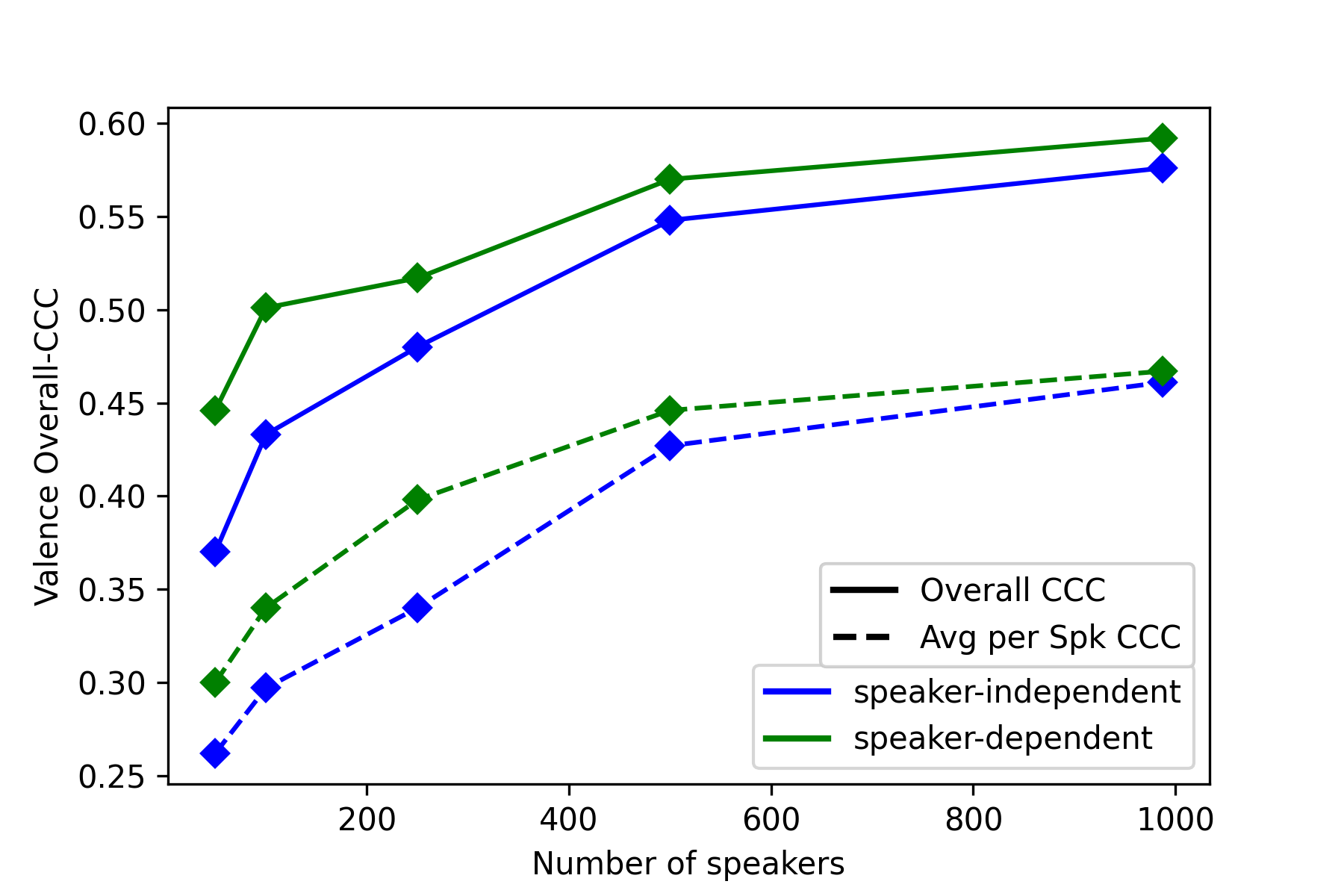}
\caption{Performance gap between speaker-dependent and speaker-independent models for valence estimation with varying the number of training speakers.}
\label{fig:per_gap}
\end{figure}

\noindent \textbf{Personalization Gap.} To motivate our proposed methodology, we investigate the potential gain from the personalization of fine-tuned HuBERT on valence regression (the dimension with the most potential gain from personalization as demonstrated by Sridhar \textit{et al.} \cite{sridhar2022unsupervised}).
In particular, we first create subsets $\mathcal{D}^k_{tr}$ of $\mathcal{D}_{tr}$ with $k$ speakers, where $k \in \{50, 100, 250, 500, 987\}$. Speaker-independent models with $k$ speakers are trained on $\mathcal{D}^k_{tr}$ sets. To make the results stable, we ensure that $\mathcal{D}^i_{tr} \subset \mathcal{D}^j_{tr}, \forall i<j$. For speaker-dependent models with $k$ speakers, we randomly remove $50$ speakers (\# speakers in \textit{test-a}) from $\mathcal{D}^k_{tr}$ to get $\hat{\mathcal{D}}^k_{tr}$, and fine-tune the models on $\hat{\mathcal{D}}^k_{tr} \cup$ \textit{test-a}. For all experiments, we fine-tune the HuBERT-base encoder on the generated sets and report the performance on \textit{test-b}. Since \textit{test-a} and \textit{test-b} share the same set of speakers, we consider the performance of speaker-dependent models to loosely correlate with the performance of \textit{supervised personalization} methods, and hence, the performance gap between speaker-dependent and speaker-independent models captures the potential gain from personalization. Figure \ref{fig:per_gap} demonstrates the inverse relationship between $k$ and the performance gap. The evaluation metric is the Concordance Correlation Coefficient (CCC $\uparrow$). It suggests that given sufficiently large and diverse training data, the pre-trained encoders become robust enough to learn both the general emotional patterns and the unique characteristics of different groups of speech expressions such that supervised training of the model on the test speakers leads to marginal gains. Hence, to enhance the performance of the pre-trained encoders for a target speaker, we can: (1) make the input data personalized (pre-processing); (2) modify the weights of the pre-trained encoder for the target speaker; or (3) adjust the label predictions to be more personalized (post-processing). Existing studies on personalized SER, \textit{e.g.}, \cite{kim2016multistage, bang2018adaptive, sridhar2022unsupervised}, focus on the first approach. In this work, we explore the other two alternatives.

\noindent \textbf{Performance variance across speakers.} Though simply fine-tuning HuBERT achieves promising overall performance on MSP-Podcast for speech emotion recognition, Wagner \textit{et al.} \cite{wagner2022dawn} find that there is a huge variance across the per-speaker performance. 
We investigate whether the performance variance is due to the feature shift or the label shift. Specifically, to measure the feature and label shift for each target speaker, we calculate the KL divergence between the feature and label distributions of the target speaker and those of the whole training set. Then we calculate the Pearson correlation coefficient (PCC) between the feature/label shift and the speaker performance. For arousal estimation, we find that the PCC between the feature shift and the regression performance is $-0.714$ while the PCC between the label shift and performance is $-0.502$. The results suggest that both feature and label shifts contribute to the performance variance. Moreover, the correlation between the feature shift and label shift is $0.285$, which suggests the potential of using features to detect and remove label shifts.

\section{Method}
\noindent \textbf{Personalized Adaptive Pre-training (PAPT).}
Inspired by prior study in task-adaptive pre-training \cite{gururangan2020don} and the problem of feature shift described above, we propose to perform adaptive pre-training on $\mathcal{D}=\{(u_i, s_i)\}^N_{i=1}$ along with trainable speaker embeddings in a self-supervised manner. Specifically, in addition to the original speech encoder $\text{E}$, we train a speaker embedding network $\text{S}$ to extract the speaker embedding $e_i = \text{S}(s_i) \in \mathbb{R}^d$,
where $d$ is the embedding size for the Transformer.
Then, the speaker embedding $e_i$ is summed with the utterance feature $f_i=\text{E}(u_i)$ to get a personalized feature representation $f^{p}_i=f_i+e_i$. For personalized pre-training, $f^{p}_i$ is used to compute the pre-training loss (cross-entropy) on pseudo-label prediction for masked frames.
\begin{equation}
\mathcal{L}_{pt}=-\sum_{i=1}^{N_b} \sum_{t=1}^{M_i} \log P(l_{it} | f^{p}_{it}),
\end{equation}
where $N_b$ is the number of utterances in the batch, $M_i$ is the number of masked frames for utterance $u_i$, and $l_{it}$ denotes the pseudo-label for the $t$-th masked frame in utterance $u_i$. For ER downstream tasks, we reduce the temporal dimension for $f^{p}_i$ by mean-pooling and feed the output to a fully-connected layer to produce the label predictions.

\noindent \textbf{Personalized Label Distribution Calibration (PLDC).} Motivated by the problem of label distribution shift described above, we further want to add a personalized post-inference technique to correct the predicted label distributions. Specifically, given the predictions for a target speaker, the main idea is to identify the most similar speakers from the train set based on the feature similarity and use their label distribution statistics (means and standard deviations) to calibrate the predicted label distributions of the target test speaker. In particular, for speaker $s$ in both the train and test set, we extract the features for each utterance of $s$ and average them to form the speaker vector
\begin{equation}
\label{eq:2}
 v_{s}=\frac{\sum_{k=1}^{N_s} \bar{\text{E}}^{p}_{ft}(u^k_{s})}{N_s},
\end{equation}
where $\text{E}^{p}_{ft}$ denotes the ER-fine-tuned model of $\text{E}^{p}$ (the personalized adapted version of $\text{E}$), $\bar{\text{E}}^{p}_{ft}(u^k_{s})$ denotes the mean-pooled vector representation for utterance $u^k_s$, and $N_s$ is the number of utterances from speaker $s$.

Then, for each speaker in the test set, we retrieve the top-k most similar speakers in the train set based on the cosine similarity between the speaker vectors. Next, we average the label distribution statistics from the retrieved speakers to get an estimation of the mean $\bar \mu$ and standard deviation $\bar \sigma$. Finally, each predicted label $y$ for the target speaker would be shifted as
\begin{equation}
 \tilde y = \frac{y - \mu}{\sigma}\times \bar \sigma + \bar \mu.
\end{equation}
where $\mu$ and $\sigma$ are the mean and standard deviation for the predicted label distribution. Optionally, if we want to only shift the mean or standard deviation, we can replace $\bar \mu$ as $\mu$ or $\bar \sigma$ as $\sigma$ in the above equation, respectively.

\section{Experiments and Discussions}
\noindent \textbf{Implementation and Training Details.} 
We perform adaptive pre-training for ten epochs using the Adam optimizer with a linear learning rate scheduler ($5\%$ warm-up and a maximum learning rate of $1e^{-5}$) on a single NVIDIA Quadro RTX8000 GPU. The models are adaptively pre-trained on the combination of the official train, validation, and \textit{test-b} sets and validated on \textit{test-a}. All other settings are identical to HuBERT's pre-training configurations. For downstream fine-tuning experiments, we add a light interpreter on top of the HuBERT encoder to process the mean-pooled extracted representations. The interpreter consists of two fully-connected layers of size $\{128,32\}$ with ReLU activation, 1D BatchNorm, and a dropout ratio of $0.1$ in-between the layers. The downstream models are fine-tuned for at most ten epochs using Adam optimizer ($5e^{-5}$ learning rate) with early stopping. Following prior work \cite{sridhar2022unsupervised}, the models are optimized with a CCC loss $\mathcal{L}_{CCC} = 1 - \text{CCC}$ for arousal and valence estimation. All of our experiments are performed with the HuBERT-large architecture, except for the personalization gap experiments, as the model used to generate the pseudo-labels for HuBERT-base is not publicly available. We report two evaluation metrics, namely the Overall CCC (\textbf{O-CCC}), which concatenates the predictions on all test speakers before computing a single CCC score for the test set, and \textbf{A-CCC}, which denotes the average CCC scores computed for each test speaker.

\noindent \textbf{Baselines.} We compare our method to three baselines: (1) Vanilla-FT in which $\text{E}$ is fine-tuned on $\mathcal{D}_{tr}$. (2) B2 represents the data weighting method proposed by Sridhar \textit{et al.} \cite{sridhar2022unsupervised}. (3) Task-Adaptive Pre-Training (TAPT) in which encoder $\text{E}$ is continued pre-training on $\mathcal{D}$ for ten epochs.

\begin{table}[t]
\footnotesize
\caption{Evaluations on MSP-Podcast (\textit{test-b}) in terms of CCC ($\uparrow$). O-CCC refers to the overall CCC between the prediction and ground truth. A-CCC denotes the average CCC for each test speaker. Numbers in the brackets are the standard deviations calculated across speakers. Our proposed PAPT-FT achieves superior performance compared to the baselines.}
\centering
\scalebox{0.9}{\begin{tabular}{l|cc|cc}
\toprule
\rowcolor{Gray}
 & \multicolumn{2}{c|}{Arousal} & \multicolumn{2}{c}{Valence} \\
\midrule
\rowcolor{Gray}
Metric & O-CCC & A-CCC & O-CCC & A-CCC \\
\midrule
Vanilla-FT & 0.712 & 0.512 & 0.607 & 0.514 \\
B2 & 0.735 & 0.517 & 0.650 & 0.569 \\
TAPT-FT & 0.717 & 0.518 & 0.630 & 0.542 \\
\midrule
PAPT-FT & \textbf{0.740} & 0.531 (0.177) & \textbf{0.663} & 0.569 (0.133) \\
+ $\mu$ shift & 0.722 & 0.528 (0.190) & 0.660 & 0.566 (0.134) \\ 
+ $\sigma$ shift & 0.732 & \textbf{0.541 (0.168)} & 0.662 & \textbf{0.578 (0.131)} \\
+ $(\mu,\sigma)$ shift & 0.713 & 0.540 (0.178) & 0.657 & 0.575 (0.131) \\
\bottomrule
\end{tabular}}
\label{tab:seen_results}
\end{table}

\noindent \textbf{Experimental Results on \textit{test-b}.} Table \ref{tab:seen_results} shows the comparison between our proposed methods and the baselines on MSP-Podcast. Compared to the best-performing baselines, our methods achieve superior performance on both arousal and valence estimation, with a gain of $0.023$ and $0.009$ on arousal and valence A-CCC respectively. Notably, we achieve state-of-the-art results for the task of valence estimation, in which our Overall-CCC score achieves $0.665$ (on the whole test set of MSP-Podcast v1.6) compared to $0.627$ as reported by Sriniva \textit{et al.} \cite{srinivasan2022representation}. When using PLDC, we can observe a significant increase in A-CCC, which suggests performance improvement for individual speakers. However, we can also see that as A-CCC improves with PLDC, O-CCC generally decreases. We attribute this to the high variance in the number of utterances of each speaker in the test set. Furthermore, Table \ref{tab:seen_results} also demonstrates that PLDC consistently achieves the best performance when we only perform $\sigma$ shifting, while $\mu$ shifting often reduces both A-CCC and O-CCC. We hypothesize that it is more difficult to estimate the mean than the (high) variance for a speaker with a wide range of arousal/valence labels.

\noindent \textbf{Evaluations on Unseen Speakers.} We further validate the robustness of our method on unseen speakers (\textit{test-c}). We directly make inference with $\text{E}^{p}_{ft}$ on \textit{test-c} without re-training any components. Specifically, for each utterance from an unseen speaker, we provide $\text{E}^{p}_{ft}$ with a training speaker embedding as a proxy for the unseen speaker. We apply the same strategy used in our PLDC module, in which we compute a vector representation for the current unseen speaker and each speaker in the train set as in Equation \ref{eq:2}. However, we use the original pre-trained encoder $\text{E}$ instead of $\text{E}^{p}_{ft}$ as the model cannot extract feature representation for the current (unseen) speaker without a proxy speaker. We then use the (seen) speaker in the train set with the highest similarity score as a proxy for the current speaker. The retrieved proxy speakers can later be used for the PLDC module to further boost prediction performance, as demonstrated in Table \ref{tab:unseen_results}. Our proposed methods outperform the baselines by a significant margin, with up to $0.030$ and $0.026$ on A-CCC for arousal and valence estimation, respectively. It is important to note that the B2 method \cite{sridhar2022unsupervised} is not applicable in this case as it would require re-adjustment of the data sample weights given the new speakers, which requires re-training the model.

\begin{table}[t]
\footnotesize
\caption{Evaluations on unseen speakers (\textit{test-c}).}
\centering
\scalebox{0.9}{\begin{tabular}{l|cc|cc}
\toprule
\rowcolor{Gray}
 & \multicolumn{2}{c|}{Arousal} & \multicolumn{2}{c}{Valence} \\
\midrule
\rowcolor{Gray}
Metric & O-CCC & A-CCC & O-CCC & A-CCC \\
\midrule
Vanilla-FT & 0.360 & 0.263 & 0.243 & 0.290 \\ 
TAPT-FT & 0.384 & 0.267 & \textbf{0.339} & 0.306 \\
\midrule
PAPT-FT & \textbf{0.398} & 0.280 & 0.321 & 0.322 \\
+ $\mu$ shift & 0.374 & 0.284 & 0.299 & 0.320 \\
+ $\sigma$ shift & 0.386 & 0.294 & 0.320 & \textbf{0.332} \\ 
+ $(\mu,\sigma)$ shift & 0.363 & \textbf{0.297} & 0.301 & 0.329 \\
\bottomrule
\end{tabular}}
\label{tab:unseen_results}
\end{table}

\begin{table}[t]
\footnotesize
\caption{Effect of different speaker embedding fusion positions.}
\centering
\scalebox{0.9}{\begin{tabular}{l|ccc|c}
\toprule
\rowcolor{Gray}
 & Last & First & Prefix & None \\
\midrule
$\mathcal{L}^{val}_{pt}$ ($\downarrow$) & \textbf{2.78} & 2.85 & 2.81 & 3.15 \\ 
A-CCC ($\uparrow$) & \textbf{0.531} & 0.519 & 0.528 & 0.512 \\
\bottomrule
\end{tabular}}
\label{tab:ablation_vec}
\end{table}

\noindent \textbf{Ablation Study.}
Table \ref{tab:ablation_vec} shows the experimental results for arousal estimation on \textit{test-b} of fine-tuned encoders (without PLDC) adaptively pre-trained with different fusion positions of the speaker embeddings. In particular, \textit{Last} refers to our proposed setting in which the speaker embeddings are added to the output of the Transformer encoder; \textit{First} refers to speaker embeddings being added to the inputs of the first layer of the Transformer encoder, and \textit{Prefix} refers to the setting in which the speaker embeddings are concatenated as prefixes to the inputs of the Transformer encoder. \textit{None} refers to the vanilla HuBERT encoder. We also provide $\mathcal{L}^{val}_{pt}$, the best pre-train loss on the validation set, \textit{i.e.}, \textit{test-a}, during the PAPT phase. 
We find that \textit{Last} provides the best results.

\section{Conclusion}
In this paper, we propose two methods to adapt pre-trained speech encoders for personalized speech emotion recognition, namely PAPT, which jointly pre-trains speech encoders with speaker embeddings to produce personalized speech representations, and PLDC, which performs distribution calibration for the predicted labels based on retrieved similar speakers. We validate the effectiveness of the proposed techniques via extensive experiments on the MSP-Podcast dataset, in which our models consistently outperform strong baselines and reach state-of-the-art performance for valence estimation. We further demonstrate the robustness of the personalized models for unseen speakers.

\section{Acknowledgement}
This work is supported by the National Science Foundation under Grant No. 2211550. Research was also sponsored by the Army Research Office and was accomplished under Cooperative Agreement Number W911NF-20-2-0053. The views and conclusions contained in this document are those of the authors and should not be interpreted as representing the official policies, either expressed or implied, of the Army Research Office or the U.S. Government. The U.S. Government is authorized to reproduce and distribute reprints for Government purposes notwithstanding any copyright notation herein.
\bibliographystyle{IEEEtran}
\bibliography{reference}

\end{document}